\documentclass[preprint,showpacs,showkeys,preprintnumbers,amsmath,amssymb]{revtex4}

\usepackage{graphicx}
\usepackage{dcolumn}

\begin{document}

\title{Exponential decay of relaxation effects at LaAlO$_3$/SrTiO$_3$
heterointerfaces}

\author{U.~Schwingenschl\"ogl and C.~Schuster}
\affiliation{Institut f\"ur Physik, Universit\"at Augsburg, 86135 Augsburg,
Germany}

\date{\today}

\begin{abstract}
We study the decay of interface induced structural and electronic
relaxation effects in epitaxial LaAlO$_3$/SrTiO$_3$ heterostructures.
The results are based on first-principles band structure calculations
for a multilayer configuration with an ultrathin LaAlO$_3$ layer sandwiched
between bulk-like SrTiO$_3$ layers. We carry out the structure optimization
for the heterointerface and investigate the electronic states of the
conducting interface layer, which is found to extend over two SrTiO$_3$ unit cells.
The decay of atomic displacements is analyzed as a function of the
distance to the interface, and the resulting exponential law is
evaluated quantitatively.
\end{abstract}

\pacs{73.20.-r, 73.20.At, 73.40.Kp}
\keywords{density functional theory, surface, interface, SrTiO$_3$, LaAlO$_3$}

\maketitle

Heterostructures based on perovskite transition metal oxides have attracted great 
attention in the last decades \cite{review1,review2}, particularly due to the discovery of extraordinary electronic and magnetic properties of the internal interfaces. For
example, a quasi two-dimensional electron gas (2DEG) with an unexpected high charge
carrier density develops at the contact between the two band insulators LaAlO$_3$
and SrTiO$_3$ \cite{ohtomo04,ohtomo06}, despite of sizeable band gaps of 5.6\,eV
and 3.2\,eV \cite{bandgap1,bandgap2}. The LaAlO$_3$/SrTiO$_3$ hetero\-interface
therefore has been subject to intensive research in recent years
\cite{LAOSTOneu1,LAOSTOneu2,us08a}. From the structure point of view it comprises
(SrO)$^0$, (TiO$_2$)$^0$, (LaO)$^+$, and (AlO$_2$)$^-$ layers, where the electron-doped
(TiO$_2$)$^0$/(LaO)$^+$ contact gives rise to the 2DEG. Moreover, the contact
between a Mott insulator and a band insulator
can likewise show conductivity, as realized in the LaTiO$_3$/SrTiO$_3$ heterostructure
\cite{LTOSTO}. In all these cases, the phenomenon of conductivity has
to be attributed to electronic relaxation, because the local charge distribution
and its alterations close to the interface are the main ingredients. Local
deviation from the bulk crystal structure as well as a transfer of charge
across the interface, induced by different electrochemical potentials in the component
materials, hence will play a key role for the formation of conduction states \cite{okamoto04}.

Electronic relaxation usually is accompanied by a fundamental lattice relaxation.
The influence of structure modifications on the LaTiO$_3$/SrTiO$_3$ interface
has been stressed by Hamann {\it et al.} \cite{hamann06}, comparing experimental data to
results from electronic structure calculations. In addition to the initial formation of the
conduction layer, the mobility of the charge carriers likewise depends
critically on structural details, such as the incorporation of O-vacancies. For the
LaAlO$_3$/SrTiO$_3$ interface, the mobility increases strongly with the vacancy concentration
\cite{herranz07}. Nevertheless, conductivity is maintained in the stoichiometric
case. Pulsed laser deposition techniques and molecular beam epitaxy nowadays, in principle,
make it possible to grow layered structures with a precision of a single unit cell,
and therefore to create atomically sharp interfaces. Local structural and electronic
properties of such contacts between two perovskite oxide compounds, in general, differ
strikingly from the bulk materials, similar to the reconstruction and formation of
specific electronic states at surfaces \cite{ge001}.

Interface atomic structures often are resolved by x-ray diffraction
techniques. A contact of LaAlO$_3$ and SrTiO$_3$ layers, for instance, is
characterized by displacements of the anions and cations in opposite direction,
because of Jahn-Teller distorted TiO$_6$ octahedra \cite{vonk07}. Data of the
Ti$^{3+}$/Ti$^{4+}$ mixture in the vicinity of the contact suggest, that we
actually have no atomically sharp interface, but a metallic
La$_{1-x}$Sr$_x$TiO$_3$ layer is formed \cite{willmott07}.
Moreover, conducting tip atomic force microscopy ensures that the metallicity is confined
to a rather thin region of $\approx$50\,nm extension \cite{basletic}. From the theoretical point of
view, first principles electronic structure calculations using density functional theory
(DFT) are a powerful tool for studying a structural reconstruction. Results have been
reported in the literature for the LaAlO$_3$/SrTiO$_3$ \cite{park06,albina07,gemming06} and
the LaTiO$_3$/SrTiO$_3$ interface \cite{hamann06,okamoto06}. An elongation of the TiO$_6$
octahedra is found to be a common feature, in accordance with experimental data. Electron
doping of Ti ions, in combination with the Jahn-Teller effect, leads to the appearance of
metallicity in a band structure calculation applying the local density approximation (LDA).
Considering the local electron-electron interaction more
accurately by means of the LDA+U scheme, Pentcheva and Pickett \cite{pentcheva06} have
succeeded in obtaining a ferromagnetic spin order, tracing back to occupied $d_{xy}$ orbitals
in a checkerboard of Ti$^{3+}$ sites.

\begin{figure}
\includegraphics[width=0.17\textwidth,clip]{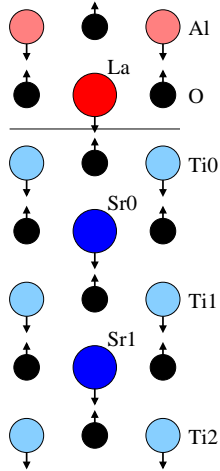}
\caption{(Color online) Schematic view of the LaAlO$_3$/SrTiO$_3$
heterostructure, in a projection along the [100] direction and
perpendicular to the interface plane. Arrows show the directions
of atomic shifts due to the interface lattice relaxation.}
\label{fig1}
\end{figure}

While structural distortions have been investigated in detail for different interfaces,
an analysis of their decay perpendicular to the contact plane is missing
so far, although a rather general behaviour is to be expected for a large class of
materials. The work of Hamann {\it et al.} \cite{hamann06} includes data about the atomic
displacements but not a systematic characterization. Furthermore, the variation of the
electrostatic potential across the LaAlO$_3$/SrTiO$_3$ interface is discussed in \cite{albina07}.
Beyond these data, investigation of structural details as a function of the distance
to the interface and, in particular, their interrelations to the electronic states is of
special interest for a quantitative description of transport processes. In order to solve
these questions, our present study deals with multilayer structures consisting of a thin
LaAlO$_3$ domain contacted to a bulk-like SrTiO$_3$ domain, which allows us to quantify the
decay of the structural and electronic relaxation effects within the titanate. We carry
out a structure optimization for the full heterointerface to give a spacially resolved
characterization of the electronic states. In particular, our data point at an exponential law
for the suppression of the interface-induced distortion, for which we evaluate
the (structural) screening length in detail. Comparison to an electrostatic model then
enables us to obtain insights into the strength of the electronic screening
in perovskite-based transition metal oxides.

Our findings rely on the generalized gradient approximation (GGA),
where we use the Wien2k code \cite{wien2k} with a mixed linear augmented-plane-wave (APW)
and APW plus local-orbitals basis set. The package is particularly suitable
for describing interfaces \cite{us07,schuster07}. In all calculations the charge
density is represented by $\approx$27{,}000 plane waves. In addition, the {\bf k}-space
grid has 21 points in the irreducible wedge of the Brillouin zone and
the Perdew-Burke-Ernzernhof parametrization is used.
The basis set for the expansion of the wave functions
contains the valence states La $6s$, $6p$, $5d$, Sr $5s$, $5p$, Ti $4s$, $4p$, $3d$, Al $3s$,
$3p$, and O $3s$, $3p$, and the semi core states La $5s$, $5p$, Sr $4s$, $4p$, Ti $4s$,
$4p$, Al $2p$, and O $2s$. We set up a tetragonal supercell by stacking
20 perovskite unit cells: A layer of two LaAlO$_3$ unit cells in the center of the supercell
is sandwiched between bulk-like SrTiO$_3$ layers, which both consist of nine unit cells.
In total, the supercell contains 42 inequivalent atomic sites.
In the $ab$-plane, i.e.\ the interface plane, and in the $c$-direction
the lattice constant is set to 3.905\,\AA, i.e.\ the value of bulk SrTiO$_3$.
We deal with an n-type interface with (TiO$_2$)$^0$/LaO$^+$ stacking.

\begin{figure}
\includegraphics[width=0.4\textwidth,clip]{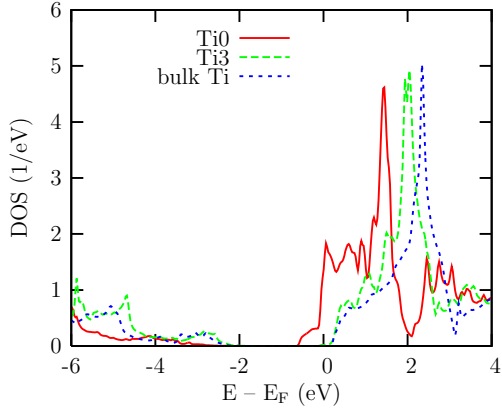}
\caption{(Color online) Partial Ti $3d$ DOS (per atom) for
the interface atoms (Ti0) and the atoms in the 4th
TiO$_2$ plane off the interface (Ti3), see Fig.\ \ref{fig1}. The
corresponding Ti $3d$ DOS of bulk SrTiO$_3$ is shown for comparison.}
\label{fig2}
\end{figure}

\begin{figure}
\includegraphics[width=0.4\textwidth,clip]{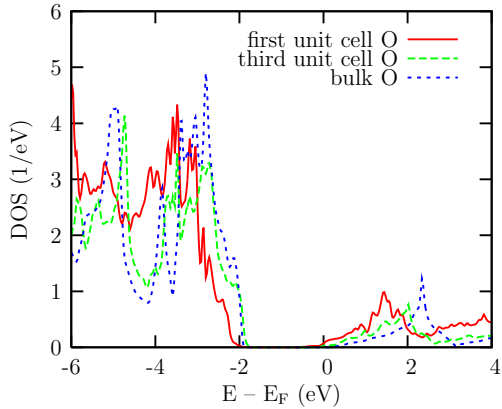}
\caption{(Color online) Comparison of partial O $2p$ DOS data (per atom) for
the O atoms in the first and third SrTiO$_3$ unit cell off the interface 
and bulk SrTiO$_3$.}
\label{fig3}
\end{figure}

The atomic structure of the LaAlO$_3$/SrTiO$_3$ supercell is illustrated schematically
in Fig.\ \ref{fig1} in a projection along the [100] axis, where the interface plane is
oriented perpendicular to the stack. For convenience, we have numbered the Ti and Sr
sites according to their distance to the interface. In Fig.\ \ref{fig2} we show the
partial Ti $3d$ DOS calculated for the Ti0 and Ti3 atoms of our heterointerface,
respectively. These data are compared to the bulk Ti $3d$ DOS. Because of a charge
transfer off the LaAlO$_3$ layer and, consequently, electron doping of the Ti atoms in
the vicinity of the interface, the Ti0 $3d$ DOS shifts to lower energy, with respect
to the bulk DOS, and conduction states are formed. As these states are characterized
by a remarkable (quasi) two-dimensional dispersion, they are attributed to a 2DEG
\cite{us08a}, in accordance with experimental findings. However, for a growing distance
to the interface the doping amplitude declines rapidly, which is visible in
Fig.\ \ref{fig2} for the Ti3 site: conduction states have disappeared. We mention
that for the neighbouring Ti2 site a small but finite DOS remains at the Fermi
energy. Moreover, since even the gross shape of the Ti3 DOS is similar to the
bulk DOS, interface effects have vanished.

Conduction states therefore are restricted to a narrow area of only two
SrTiO$_3$ unit cells, i.e.\ three TiO$_2$ planes. However, due to electronic
correlations beyond the GGA which tend to increase band gaps, this restriction
even is only an upper boundary for the extension of the electron gas.
Our line of reasoning is supported by the O $2p$ DOS depicted in Fig.\ \ref{fig3}.
The data again refer to the first and third SrTiO$_3$ unit cell off the interface and
are compared to the bulk DOS. While close to the LaAlO$_3$ contact the O electronic
structure reveals serious deviations from the bulk state, only minor differences are left
for the O sites in the third unit cell.

\begin{figure}
\includegraphics[width=0.23\textwidth,clip]{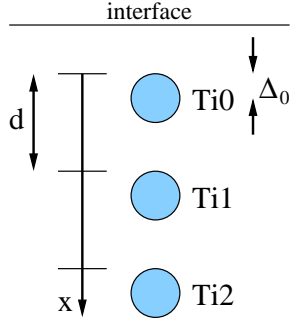}
\caption{(Color online) Setup of the electrostatic model.}
\label{figmod}
\end{figure}
The dependence of the structure relaxation on the distance of an
atom from the
interface can be described by a merely electrostatic model.
Assuming that all the atomic shifts are directed perpendicular to
the contact plane, we decompose the system into Ti, Sr, and O chains
running in this direction. The equilibrium position
$\Delta=0$ of any atom then is determined by the minimum of the
electrostatic potential $u(\Delta)$ due to its nearest neighbours,
\begin{displaymath}
4\pi\epsilon_0u(\Delta)=(d+\Delta)^{-1}+(d-\Delta)^{-1},
\end{displaymath} 
where $d$ denotes the nearest neighbour distance. In harmonic approximation,
the atomic force as function of the displacement from equilibrium therefore is
\begin{displaymath}
F(\Delta)=-\frac{4}{4\pi\epsilon_0d^3}\;\Delta.
\end{displaymath}
Let $x$ denote the distance of an atom from the first
atom of its chain at the interface, see Fig.\ \ref{figmod}, which is subject to the
initial displacement $\Delta(x=0)=\Delta_0$. We obtain
\begin{displaymath}
-\frac{4}{d^3}\;\Delta(x)=F_+(x)-F_-(x),
\end{displaymath}
where $F_\pm(x)=d^{-2}-(d-\Delta(x\mp d))^{-2}$ represents the forces of the two
neighbours acting on an atom near the interface.
A Taylor expansion up to first order gives
\begin{displaymath}
-4\Delta(x)=2\Delta(x+d)-2\Delta(x-d)=4d\Delta'(x),
\end{displaymath}
and hence $\Delta(x)=\Delta_0\exp(-d^{-1}x)$.
Electronic screening is expected to modify the exponential decay
according to
\begin{displaymath}
\Delta(x)=\Delta_0\exp(-\lambda\cdot x),
\end{displaymath}
where we set $d=1$ and introduce an effective structural screening length $\lambda>1$.

\begin{figure}
\includegraphics[width=0.4\textwidth,clip]{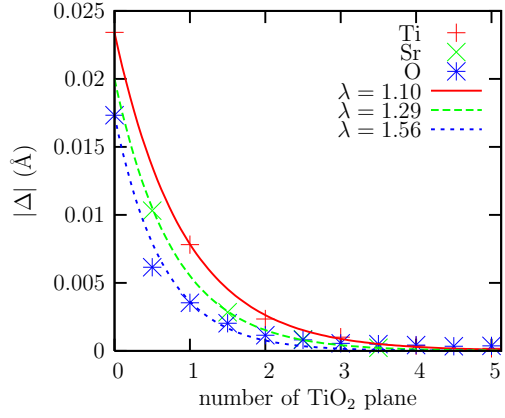}
\caption{(Color online) Calculated displacements in the fully
relaxed heterostructure given as a function of the distance to the interface
TiO$_2$ plane. The numbering of the TiO$_2$ planes is illustrated
in Fig.\ \ref{fig1}.}
\label{fig4}
\end{figure}
\begin{figure}
\includegraphics[width=0.4\textwidth,clip]{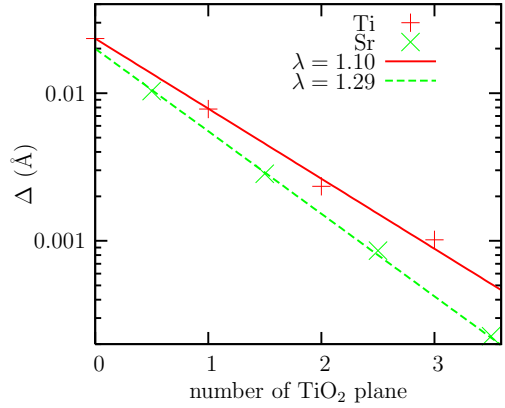}
\caption{(Color online) Logarithmic plot of the data from Fig.\ \ref{fig4}.
A nearly ideal exponential dependence on the distance to the interface
is obtained for the magnitudes of the Ti and Sr atomic displacements.}
\label{fig5}
\end{figure}

The Ti and O atomic displacements found in our heterostructure after the structure optimization
are shown in Fig.\ \ref{fig4} as functions of the distance to the interface, in each case
given via the number of the TiO$_2$ plane a site belongs to. Moreover, Fig.\ \ref{fig4}
includes the displacements of the interstitial Sr atoms. The amplitudes of all these
shifts are largely consistent with our model. Specifically, we observe a nearly perfect
exponential decay of the displacements with structural screening lengths of $\lambda=1.10$
for the Ti, $\lambda=1.29$ for the Sr, and $\lambda=1.56$ for the O sites.
Corresponding fit curves are shown in Figs.\ \ref{fig4} and \ref{fig5}, where
the logarithmic representation of the curves in Fig.\ \ref{fig5} stresses that
deviations from the exponential law are negligible for both Ti and Sr. They are
slightly larger for the O sites. The full deformation pattern induced by the
interface lattice relaxation is shown in Fig.\ \ref{fig1}. As the Ti and O
atoms are shifted in opposite direction, the experimental Jahn-Teller distortion
of the TiO$_6$ octahedra at the interface is reproduced by our data \cite{vonk07}.

The maximal amplitude of the displacements amounts to $\Delta_0=0.023$\,\AA\ for the Ti and
$\Delta_0=0.017$\,\AA\ for the O atoms, as realized for the interface TiO$_2$ layer.
Furthermore, extrapolation of the Sr data to this layer leads to a value of $\Delta_0=0.020$\,\AA,
which seems to reflect a similar interface effect on the different atomic
species. However, whereas the Ti and Sr sites move off
the interface plane, the O sites approach it. For this reason, we have
a relative shift between the Ti/Sr and the O sublattice, which is (exponentially) suppressed
in the SrTiO$_3$ bulk. Consequently, the structural prerequisites for the formation
of a 2DEG are fulfilled only in the vicinity of the interface plane. In the third
TiO$_2$ layer off the contact relaxation effects have decayed to less than 4\%
of their initial amplitude, which explains our previous observation that no conduction
states are formed here as well as in any layer farther away from the contact. Deviations
of the structural screening length from the value $\lambda=1$, as expected in the
electrostatic picture, amount to 10\% and 29\% for the Ti and Sr sites, respectively.
They can be attributed to the electronic screening, whereas the quicker decay of the O
displacements is connected to the potential of the O octahedra to evade mechanical
stress via distortions.

In conclusion, we have presented first-principles band structure results for a
prototypical LaAlO$_3$/SrTiO$_3$ heterointerface. In general, the electronic properties of
heterostructures are closely related to the structural distortions
affecting the component materials. By means of a detailed structure optimization
we have investigated the length scala on which such distortions are suppressed in
the bulk materials. We find that the decay of structural relaxation effects is
well described by an exponential law, in agreement with electrostatic
considerations. However, the structural screening length $\lambda$ is significantly reduced
due to electronic screening. We expect that these results are fairly independent
of the compounds forming the interface, as specialties of the chemical
bonding are found to play a less important role for the structural screening. In the
LaAlO$_3$/SrTiO$_3$ heterostructure the formation of a metallic layer is restricted
to a narrow area next to the interface, comprising only two perovskite unit cells.

\subsection*{Acknowledgement}
We thank U.\ Eckern, T.\ Kopp, and J.\ Mannhart for helpful discussions, and
the Deutsche Forschungsgemeinschaft for financial support (SFB 484).

\end{document}